\documentclass[12pt,prd,aps,amssymb,amsmath,tightenlines,showpacs,a4paper]{revtex4}
\def\half{\textstyle{\frac{1}{2}}}
\def\sixth{\textstyle{\frac{1}{6}}}
\def\vf{\varphi}
\def\ve{\varepsilon}
\def\p{\partial}
\def\pslash{\p\raise.3ex \hbox{\kern-.5em /}}
\def\delslash{\nabla\raise.3ex \hbox{\kern-.7em /}}
\def\beq{\begin{equation}}
\def\eeq{\end{equation}}
\def\bea{\begin{eqnarray}}
\def\eea{\end{eqnarray}}
\begin{document}

\title{Dual $\cal PT$-Symmetric Quantum Field Theories}

\author{Carl~M.~Bender${}^1$}\email{cmb@wustl.edu}
\author{H.~F.~Jones${}^2$}\email{h.f.jones@imperial.ac.uk}
\author{R.~J.~Rivers${}^2$}\email{r.rivers@imperial.ac.uk}

\affiliation{${}^1$Department of Physics, Washington University, St. Louis, MO
63130, USA \\
${}^2$Physics Department, Imperial College, London SW7 2BZ, UK}

%\date{\today}

\begin{abstract}
Some quantum field theories described by non-Hermitian Hamiltonians
are investigated. It is shown that for the case of a free fermion field theory
with a $\gamma_5$ mass term the Hamiltonian is $\cal PT$-symmetric. Depending
on the mass parameter this symmetry may be either broken or unbroken. When the
$\cal PT$ symmetry is unbroken, the spectrum of the quantum field theory is
real. For the $\cal PT$-symmetric version of the massive Thirring model in
two-dimensional space-time, which is dual to the $\cal PT$-symmetric
scalar Sine-Gordon model, an exact construction of the $\cal C$ operator is
given. It is shown that the $\cal PT$-symmetric massive Thirring and Sine-Gordon
models are equivalent to the conventional Hermitian massive Thirring and
Sine-Gordon models with appropriately shifted masses.
\end{abstract}

\maketitle
\section{Introduction}
\label{s1}
Hermiticity is a convenient symmetry of a quantum-mechanical Hamiltonian $H$
because it guarantees that the energy eigenvalues of $H$ are real. (We use the
term {\it Hermiticity} here in the usual Dirac sense, where Dirac conjugation
$H^\dagger$ represents the combined transpose and complex conjugate of $H$.)
However, there are also non-Hermitian Hamiltonians whose eigenvalues are real. A
recently discovered class of such Hamiltonians is \cite{BB}
\beq
H=p^2+x^2(ix)^\epsilon\quad(\epsilon>0).
\label{e1}
\eeq
The eigenvalues of $H$ in (\ref{e1}) are entirely real because $H$ is invariant
under space-time reflection and this reflection symmetry, known as $\cal PT$
{\it symmetry}, is unbroken \cite{BBJ,DDT}. We say that a Hamiltonian $H$ is
$\cal PT$-symmetric if $H$ commutes with the $\cal PT$ operator, where $\cal P$
represents parity reflection and $\cal T$ represents time reversal.

It is necessary that the eigenvalues of a Hamiltonian $H$ be real in order for
$H$ to describe a physical theory of quantum mechanics, but the reality of the
spectrum of $H$ is not sufficient. One must also establish that the time
evolution operator $U=e^{iHt}$ is unitary (norm-preserving). If $H$ is
non-Hermitian, then $U$ is not unitary with respect to the standard Hilbert
space inner product definition $\langle A|B\rangle=A^\dagger\cdot B$. However,
Ref.~\cite{BBJ} demonstrated that there is a new positive inner product $\langle
A|B\rangle=A^{\cal CPT}\cdot B$ with respect to which the time evolution
operator $U$ is unitary. The operator $\cal C$ is linear and is defined by the
following system of three algebraic equations:
\bea
{\cal C}^2&=&1, \label{e2}\\
\left[{\cal C},{\cal PT}\right]&=&0, \label{e3}\\
\left[{\cal C},H\right]&=&0.  \label{e4}
\eea
By solving these three simultaneous equations for the operator $\cal C$, one
obtains an inner product with respect to which $H$ is self-adjoint.

In Refs.~\cite{BBJ2,BBJ3} it is shown that the $\cal C$ operator has the
general form
\beq
{\cal C}=e^Q{\cal P},
\label{e5}
\eeq
where $Q$ is a Hermitian operator. This result makes contact with the general
framework of pseudo-Hermitian Hamiltonians \cite{AM}. In Ref.~\cite{AM} it is
shown that the square root of the positive operator $\eta\equiv e^{-Q}$ can be
used to construct a Hermitian Hamiltonian $h$ that corresponds to the
non-Hermitian Hamiltonian $H$. The Hamiltonians $h$ and $H$ are related by the
similarity transformation
\beq
h=e^{-Q/2}He^{Q/2}.
\label{e6}
\eeq

Perturbative methods were adopted in Refs.~\cite{BBJ2,BBJ3,BMW} for calculating
the $\cal C$ operator in (\ref{e5}) for the case of the cubic $\cal P
T$-symmetric quantum-mechanical Hamiltonian
\beq
H=\half p^2+\half x^2+i\ve x^3
\label{e7}
\eeq
and for the cubic $\cal PT$-symmetric field-theoretic Hamiltonian density
\beq
{\cal H}=\half\pi^2+\half(\nabla\vf)^2+\half m^2\vf^2+i\ve\vf^3.
\label{e8}
\eeq
In addition, the $\cal C$ operator has been obtained perturbatively for the
quantum-field-theoretic Lee model \cite{BBCW} and for a $\cal PT$-symmetric
version of quantum electrodynamics \cite{BCMS}. Semiclassical methods have also
been used to construct $\cal C$ \cite{BJ}.

It is explained in Refs.~\cite{BBJ2,BBJ3} how to calculate the $\cal C$ operator
using perturbative methods. To summarize the procedure, we write the Hamiltonian
$H$ in the form
\beq
H=H_0+\ve H_1,
\label{e9}
\eeq
where $H_0$ and $H_1$ are respectively even and odd under parity reflection. We
then represent the $Q$ operator as a formal series in powers of $\ve$:
\beq
Q=\ve Q_1+\ve^3Q_3+\ve^5Q_5+\cdots.
\label{e10}
\eeq
It has been shown that only odd powers of $\ve$ appear in this series
\cite{BBJ3}. Substituting (\ref{e9}) and (\ref{e10}) into (\ref{e4}) and
collecting powers of $\ve$, we obtain a sequence of algebraic equations that can
be solved successively to determine the coefficients $Q_1$, $Q_3$, $Q_5$, and so
on. The first three of these equations are
\bea
[Q_1,H_0]&=&2H_1,\cr 
[Q_3,H_0]&=&\sixth[Q_1,[Q_1,H_1]],\cr
[Q_5,H_0]&=&-\textstyle{\frac{1}{360}}[Q_1,[Q_1,[Q_1,[Q_1,H_1]]]]+\sixth[Q_1,[
Q_3,H_1]]+\sixth[Q_3,[Q_1, H_1]].
\label{e11}
\eea
In principle, the system of equations (\ref{e11}) can be solved for the
perturbation coefficients $Q_1$, $Q_3$, $Q_5$, $\cdots$ to any finite order.
However, for the cubic quantum theories that have been studied using this
procedure, it has not so far been possible to sum the perturbation series in
(\ref{e10}) to all orders.

For the field-theoretic models considered in the present paper, we are able
to calculate the perturbation coefficients $Q_1$, $Q_3$, $Q_5$, $\cdots$ to all
orders and subsequently to sum the series (\ref{e10}) {\it exactly}. Using this
procedure we can construct the $Q$ operator in closed form for a $\cal P
T$-symmetric version of the massive Thirring model in $(1+1)$-dimensional
space-time. This procedure can be formally extended to the Thirring model in
$(3+1)$ dimensions, even though this model is not renormalizable in higher
dimensions. The special feature of $(1+1)$ dimensions is that bosonization
prescriptions exist that permit us to construct bosonic theories dual to these
fermionic theories. In the present case the massive Thirring model is dual to
the purely bosonic Sine-Gordon model. It then follows that $Q$ is immediately
calculable for the dual bosonic theory.

This paper is organized as follows: In Sec.~\ref{s2} we examine a $\cal P
T$-symmetric free fermionic field theory with a $\gamma_5$ mass term and show
that there is a region of unbroken $\cal PT$ symmetry for which the energy
levels are real and a region of broken $\cal PT$ symmetry for which the energy
levels are complex. We calculate the exact $Q$ operator for this theory. In
Sec.~\ref{s3} we study the $\cal PT$-symmetric version of the massive Thirring
model in $(1+1)$-dimensional space-time and in Sec.~\ref{s4} we study the
equivalent bosonic $\cal PT$-symmetric Sine-Gordon model in $(1+1)$-dimensional
space-time. For these models we calculate the $Q$ operator exactly and in
closed form and identify the corresponding Hermitian Hamiltonians.

\section{$\cal PT$-Symmetric Free Fermion Theory with a $\gamma_5$ Mass Term}
\label{s2}

The Lagrangian density for a conventional Hermitian free fermion field theory is
\beq
{\cal L}({\bf x},t)=\bar\psi({\bf x},t)(i\pslash-m)\psi({\bf x},t)
\label{e12}
\eeq
and the corresponding Hamiltonian density is
\beq
{\cal H}({\bf x},t)=\bar\psi({\bf x},t)(-i\delslash+m)\psi({\bf x},t),
\label{e13}
\eeq
where $\bar\psi({\bf x},t)=\psi^\dagger({\bf x},t)\gamma_0$.

\subsection{Free Fermion Theories in Two-Dimensional Space-Time}

We consider first the case of two-dimensional space-time. In $(1+1)$-dimensional
space-time we adopt the conventions used in Ref.~\cite{2DQFT}:
\beq
\gamma_0=\left(\begin{array}{cc}0 & 1\\ 1 & 0 \end{array}\right)
\quad{\rm and}\quad
\gamma_1=\left(\begin{array}{cc}0 & 1\\ -1 & 0 \end{array}\right).
\label{e14}
\eeq
With these definitions we have $\gamma_0^2=1$ and $\gamma_1^2=-1$. We also
define
\beq
\gamma_5=\gamma_0\gamma_1=\left(\begin{array}{cc}1&0\\0&-1\end{array}\right),
\label{e15}
\eeq
so that $\gamma_5^2=1$.

In $(1+1)$ dimensions the parity-reflection operator $\cal P$ has the effect
\bea
{\cal P}\psi(x,t){\cal P}&=&\gamma_0\psi(-x,t),\nonumber\\
{\cal P}\bar\psi(x,t){\cal P}&=&\bar\psi(-x,t)\gamma_0.
\label{e16}
\eea
The effect of the time-reversal operator $\cal T$ is similar to that of the
parity operator,
\bea
{\cal T}\psi(x,t){\cal T}&=&\gamma_0\psi(x,-t),\nonumber\\
{\cal T}\bar\psi(x,t){\cal T}&=&\bar\psi(x,-t)\gamma_0,
\label{e17}
\eea
except that $\cal T$ is {\it anti}-linear and therefore takes the complex
conjugate of complex numbers.

It is easy to see that the Hamiltonian $H=\int dx\,{\cal H}(x,t)$, where $\cal
H$ is given in (\ref{e13}), is Hermitian: $H=H^\dagger$. Also, $H$ is separately
invariant under parity reflection and under time reversal:
$${\cal P}H{\cal P}=H\quad{\rm and}\quad {\cal T}H{\cal T}=H.$$

\begin{figure}[b!]\vspace{2.5in}
\includegraphics{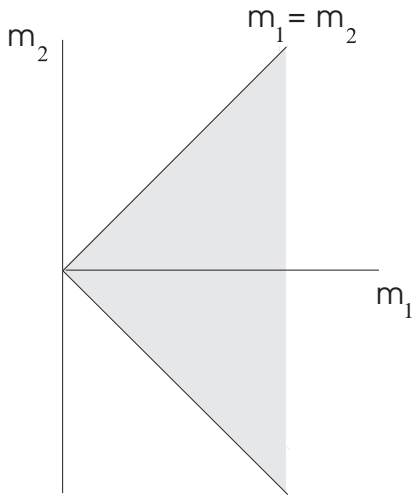}
\caption{Parametric regions of broken and unbroken $\cal PT$ symmetry for the
Hamiltonian $H$ in (\ref{e18}) in the $(m_1,m_2)$ plane. The region of unbroken
$\cal PT$ symmetry $m_1^2\geq m_2^2$ is shaded. For these values of the
parameters $m_1$ and $m_2$, the Dirac equation describes the propagation of
particles having real mass. The special case of Hermiticity is obtained on the
line $m_2=0$, which lies at the center of the region of unbroken $\cal PT$
symmetry. The region of broken $\cal PT$ symmetry $m_1^2< m_2^2$ is unshaded.}
\label{f1}
\end{figure}

Now let us construct a {\it non}-Hermitian Hamiltonian by adding a
$\gamma_5$-dependent mass term to the Hamiltonian density in (\ref{e13}):
\beq
{\cal H}(x,t)=\bar\psi(x,t)(-i\delslash+m_1+m_2\gamma_5)\psi(x,t)\quad(m_2~{\rm
real}).
\label{e18}
\eeq
The Hamiltonian $H=\int dx\,{\cal H}(x,t)$ associated with this Hamiltonian
density is not Hermitian because the $m_2$ term changes sign under Hermitian
conjugation. This sign change occurs because $\gamma_0$ and $\gamma_5$
anticommute. Also, $H$ is not invariant under $\cal P$ or under $\cal T$
separately because the $m_2$ term changes sign under each of these reflections.
However, $H$ {\it is} invariant under combined $\cal P$ and $\cal T$ reflection.
Thus, $H$ is $\cal PT$-symmetric: $H^{\cal PT}={\cal PT}H{\cal PT}=H.$ The
field equation associated with $\cal H$ in (\ref{e18}) is 
\beq
\left(i\pslash-m_1-m_2\gamma_5\right)\psi(x,t)=0.
\label{e19}
\eeq
If we iterate this equation and use $\pslash^2=\partial^2$, we obtain 
the two-dimensional Klein-Gordon equation
\beq
\left(\partial^2+\mu^2\right)\psi(x,t)=0,
\label{e20}
\eeq
where $\mu^2=m_1^2-m_2^2$. Thus, the physical mass that propagates under this
equation is real when the inequality
\beq
m_1^2\geq m_2^2
\label{e21}
\eeq
is satisfied. This condition defines the two-dimensional parametric region of
{\it unbroken} $\cal PT$ symmetry. When (\ref{e21}) is not satisfied, the $\cal
PT$ symmetry is {\it broken}. Figure~\ref{f1} displays the regions of broken and
unbroken $\cal PT$ symmetry. Note that the special case of Hermiticity is 
restricted to a one-dimensional region (the line $m_2=0$).

We will now show how to calculate the $\cal C$ operator associated with the
$\cal PT$-symmetric Hamiltonian density $\cal H$ in (\ref{e18}). We begin by
letting $m_1=m$ and $m_2=\ve m$ and rewriting $\cal H$ in the form
\beq
{\cal H}(x,t)=\bar\psi(x,t)(-i\delslash+m+\ve m\gamma_5)\psi(x,t).
\label{e22}
\eeq
We treat $\ve$ as a perturbation parameter and decompose the Hamiltonian
associated with $\cal H$ in (\ref{e22}) as in (\ref{e9}): $H_0=\int\!\!\int\!dx
\,dy\,\bar\psi(x,t)D(x,y)\psi(y,t)$, where $D(x,y)=(-i\delslash+m)\delta(x-y)$
and $H_1=\int\!dx\,\bar\psi(x,t)m\gamma_5\psi(x,t)$.

We seek the $\cal C$ operator in the form (\ref{e5}). Note that because $\cal C$
commutes with the Hamiltonian, it is time independent. Thus, we may calculate
the $Q$ operator at any time. For the Hermitian Hamiltonian $H_0$ the $Q$
operator vanishes. However, on the basis of the work done in Ref.~\cite{BBJ2} we
expect that when $\ve\neq0$, $Q$ takes the form of the perturbation expansion in
(\ref{e10}). We assume that the $Q_n$ operator is bilinear in the fields:
\beq
Q_n=\int\!\!\!\int \!dx\,dy\,\bar\psi(x,t)G_n(x,y)\psi(y,t),
\label{e23}
\eeq
where $G_n(x,y)$ are functions to be determined. Then, the first of the
commutation relations in (\ref{e11}) reads
\bea
\int\!\!\!\int\!\!\!\int\!\!\!\int\!dx\,dy\,dz\,dw\,[\bar\psi(x,t)G_1(x,y)
\psi(y,t),\bar\psi(z,t)D(z,w)\psi(w,t)]\cr\hspace{3cm}
=2m\!\int\! dx\,\bar\psi(x,t)\gamma_5\psi(x,t).
\label{e24}
\eea

The canonical equal-time anticommutation relations $\{\psi^\dag(x,t),\psi(y,t)\}
=\delta(x-y)$ may be used to simplify the left side of this equation:
\bea
-\int\!\!\!\int\! dx\,dy\,\bar\psi(x,t)\left(m[\gamma_0,G_1(x,y)]+i\{\gamma_0
\delslash,G_1(x,y)\}\right)\psi(y,t)\cr\hspace{2.5cm}
=2m\!\int\!dx\,\bar\psi(x,t)\gamma_5\psi(x,t).
\label{e25}
\eea
This equation does not determine $G_1(x,y)$ uniquely,
but $G_1(x,y)=-\gamma_1\delta(x-y)$
is a particular solution \cite{SOL}. Inserting this solution into the second
equation of (\ref{e11}), we find that the right side reduces to $(2/3)H_1$,
giving $G_3=G_1/3$. The third equation of (\ref{e11}) gives $G_5=G_1/5$. 

This process successively generates a hyperbolic arc tangent, with the 
all-orders result
\beq
Q=-\tanh^{-1}\!\ve\!\int\!dx\,\bar\psi(x,t)\gamma_1\psi(x,t)=-\tanh^{-1}\!\ve
\!\int\! dx\,\psi^\dag(x,t)\gamma_5\psi(x,t).
\label{e26}
\eeq
(The inverse hyperbolic tangent function in this equation requires that $|\ve|
\leq1$, or equivalently $m_1^2\geq m_2^2$, which corresponds to the shaded
region of unbroken $\cal PT$ symmetry in Fig.~\ref{f1}.) We can verify
(\ref{e26}) by constructing
\beq
h=\exp\left[\half\tanh^{-1}\!\ve\int\!dx\,\psi^\dag(x,t)\gamma_5\psi(x,t)
\right]H\exp\left[-\half\tanh^{-1}\!\ve\int\!dx\,\psi^\dag(x,t)\gamma_5\psi
(x,t)\right].
\label{e27}
\eeq
By virtue of the Lorentz-like commutation relations
\bea
[\gamma_5,\gamma_0]=-2\gamma_1,\qquad [\gamma_5,\gamma_1]=-2\gamma_0,
\label{e28}
\eea
$h$ in (\ref{e27}) reduces to
\beq
h=\!\int\!dx\,\bar\psi(x,t)(-i\delslash+\mu)\psi(x,t),
\label{e29}
\eeq
where
\beq
\mu^2=m^2(1-\ve^2)=m_1^2-m_2^2,
\label{e30}
\eeq
in agreement with (\ref{e20}). The only effect in going from $H$ to $h$ is to
change the $\gamma_5$-dependent mass term $m\bar\psi(1+\ve\gamma_5)\psi$ to a
normal mass term $\mu\bar\psi\psi$. We conclude that the non-Hermitian $\cal P
T$-symmetric Hamiltonian density in (\ref{e18}) is equivalent to the Hermitian
Hamiltonian density in (\ref{e13}) with $m$ replaced by $\mu$.

\subsection{Free Fermion Theories in Four-Dimensional Space-Time}

In $(3+1)$ dimensions the analogs of (\ref{e16}) and (\ref{e17}) are
\bea
{\cal P}\psi(\mathbf{x},t){\cal P}&=& \gamma_0\psi(-\mathbf{x},t),\cr
{\cal P}\bar\psi(\mathbf{x},t){\cal P}&=&\bar\psi(-\mathbf{x},t)\gamma_0,
\label{e31}
\eea
and
\bea
{\cal T}\psi(\mathbf{x},t){\cal T}&=&{\bf C}^{-1}\gamma_5\psi(\mathbf{x},-t),\cr
{\cal T}\bar\psi(\mathbf{x},t){\cal T}&=&\bar\psi(\mathbf{x},-t)\gamma_5{\bf C},
\label{e32}
\eea
where $\bf C$ is the charge-conjugation matrix, defined by ${\bf C}^{-1}
\gamma_\mu {\bf C}=-\gamma_\mu^{\rm T}$. In $(1+1)$
dimensions $\bf C$ reduces to $\gamma_0$, as in Eq.~(\ref{e17}).

The resulting $Q$ is given by
\beq
Q=-\tanh^{-1}\!\ve\!\int\!d{\bf x}\,\psi^\dag({\bf x},t)\gamma_5\psi({\bf x},t),
\label{e33}
\eeq
which is the three-dimensional generalization of the right side of (\ref{e26}).
The validity of (\ref{e33}) is due to the commutation relations
\beq
[\gamma_5,\gamma_0]=-2\gamma_0\gamma_5\quad
[\gamma_5,\gamma_0\gamma_5]=-2\gamma_0,
\label{e34}
\eeq
which are the generalizations of (\ref{e28}). Moreover, the invariance
of the kinetic term is assured by the commutation relation $[\gamma_5,
\gamma_0\gamma_\mu]=0$.

In $(3+1)$ dimensions the Hermitian Hamiltonian $h$, which is equivalent to the
non-Hermitian Hamiltonian $H$ with a $\gamma_5$ mass term, again has the shifted
mass $\mu$ in (\ref{e30}).

\section{$\cal PT$-Symmetric Massive Thirring Model}
\label{s3}

Our starting point, again in (1+1) dimensions, is the Lagrangian density for the
conventional massive Thirring model
\beq
{\cal L}=\bar\psi(i\pslash-m)\psi+\half g(\bar\psi\gamma^\mu\psi)(\bar\psi
\gamma_\mu\psi),
\label{e35}
\eeq
with the corresponding Hamiltonian density
\beq
{\cal H}=\bar\psi(-i\delslash+m)\psi-\half g(\bar\psi\gamma^\mu\psi)(\bar\psi
\gamma_\mu\psi),
\label{e36}
\eeq
This model is known \cite{2DQFT} to be equivalent to the Sine-Gordon model
(see Sec.~\ref{s4}) with the correspondence
\beq
\frac{\lambda^2}{4\pi}=\frac{1}{1-g/\pi},
\label{e37}
\eeq
so that, in particular, the free fermion theory is equivalent to the Sine-Gordon
model with the special value for the coupling constant $\lambda^2=4\pi$.

The modification analogous to that of Eq.~(\ref{e22}) is the introduction of a
$\gamma_5$-dependent mass according to
\beq
{\cal H}=\bar\psi(-i\delslash+m+\ve m\gamma_5)\psi-\half g(\bar\psi\gamma^\mu
\psi)(\bar\psi\gamma_\mu\psi),
\label{e38}
\eeq
The additional term is non-Hermitian but $\cal PT$-symmetric because it is odd
under both parity reflection and time reversal.

In Sec.~\ref{s2} we considered the case $g=0$. The $Q$ operator for the
interacting case $g\neq0$ is in fact identical to the $Q$ operator for the case
$g=0$ because in $(1+1)$-dimensional space the interaction term $(\bar\psi\gamma
^\mu\psi)(\bar\psi\gamma_\mu\psi)$ commutes with the $Q$ in (\ref{e26}). Thus,
once again we conclude that the non-Hermitian $\cal PT$-symmetric Hamiltonian
density in (\ref{e38}) is equivalent to the Hermitian Hamiltonian density in
(\ref{e36}) with the mass $m$ replaced by $\mu$ in (\ref{e30}).

The same holds true for the $(3+1)$-dimensional interacting Thirring model by
virtue of the commutation relation $[\gamma_5,\gamma_0\gamma_\mu]=0$, but
because this higher-dimensional field theory is nonrenormalizable, the $Q$
operator may only have a formal significance.

\section{$\cal PT$-Symmetric Sine-Gordon Model}
\label{s4}

The massive Thirring Model (\ref{e35}) is dual to the conventional Sine-Gordon
model in (1+1) dimensions whose Lagrangian density is
\beq
{\cal L}=\half(\p_\mu\vf)^2+\frac{m^2}{\lambda^2}(\cos\lambda\vf-1),
\label{e39}
\eeq
and whose corresponding Hamiltonian density is
\beq
{\cal H}=\half\pi^2+\half(\nabla\vf)^2+\frac{m^2}{\lambda^2}(1-\cos\lambda\vf),
\label{e40}
\eeq
where $\pi(x,t)=\partial_0\vf(x,t)$, and in $(1+1)$-dimensional space $\nabla\vf
(x,t)$ is just $\p_1\vf(x,t)$.

The $\cal PT$-symmetric extension  (\ref{e38}) of the modified Thirring model
is, by the same analysis, dual to a modified Sine-Gordon model with
Hamiltonian density
\beq
{\cal H}=\half\pi^2+\half(\nabla\vf)^2+\frac{m^2}{\lambda^2}(1-\cos\lambda\vf-i
\ve\sin\lambda\vf),
\label{e41}
\eeq
which is $PT$-symmetric but no longer Hermitian.

On the basis of duality, the form of the operator $Q_1$ is easy to guess:
\beq
Q_1=\xi_1\!\int\!dx\,\pi(x,t),
\label{e42}
\eeq
where $\xi_1$ is a constant. The commutator of $Q_1$ with $H_0$ is
\bea
\left[Q_1, H_0\right]&=&\xi_1\!\int\!\!\!\int\! dx\,dy\,\left[\pi(y,t),\half
[\nabla\vf(x,t)]^2-\frac{m^2}{\lambda^2}\cos[\lambda\vf(x,t)]\right]\cr &&\cr
&=&i\xi_1\!\int\!dx\,\left(\nabla^2\vf(x,t)-\frac{m^2}{\lambda}\sin[
\lambda\vf(x,t)]\right),
\label{e43}
\eea
by virtue of the canonical commutation relation $[\vf(x,t),\pi(y,t)]=i\delta
(x-y)$. The first term, being a total derivative, integrates to zero, and
identifying the remainder with $2H_1$ gives $\xi_1=2/\lambda$.

The form of $Q_3$ is the same as that for $Q_1$ with $\xi_1$ replaced by the
constant $\xi_3$. Substituting into the second equation of (\ref{e11}), we
obtain $\xi_3=\xi_1/3$. Similarly we obtain $\xi_5=\xi_1/5$, giving a strong
indication that we are again generating the series for $\tanh^{-1}\!\ve$ and
that the all-orders result for $Q$ is
\beq
Q=\frac{2\tanh^{-1}\!\ve}{\lambda}\!\int\!dx\,\pi(x,t).
\label{e44}
\eeq
We can verify this result {\it a posteriori} by constructing the equivalent
Hermitian Hamiltonian:
% $h$:
\beq
h=\exp\left[-\frac{\tanh^{-1}\!\ve}{\lambda}\!\int\!dx\,\pi(x,t)\right]H
\exp\left[\frac{\tanh^{-1}\!\ve}{\lambda}\!\int\!dx\,\pi(x,t)\right].
\label{e45}
\eeq

It is interesting that the operation that transforms $H$ to $h$ has precisely
the effect of shifting the boson field $\vf$ by an imaginary constant:
\bea
\vf\to\vf+i\frac{\tanh^{-1}\!\ve}{\lambda}.
\label{e46}
\eea
Under this transformation the interaction term $m^2\lambda^{-2}(1-\cos\lambda\vf
-i\ve\sin\lambda\vf)$ in (\ref{e41}) becomes $-m^2\lambda^{-2}(1-\ve^2)\cos
\lambda\vf$, apart from an additive constant. Hence, $h$ is the Hamiltonian for
the conventional Sine-Gordon model, but with mass $\mu$ given by (\ref{e30}).
This change in the mass is exactly the same as we observed in the fermionic
theory discussed in Sec.~\ref{s3}. Note that $h$, being Hermitian, is even in
the parameter $\ve$ that breaks the Hermiticity of $H$.

The idea of generating a non-Hermitian but $\cal PT$-symmetric Hamiltonian from
a Hermitian Hamiltonian by shifting the field operator as in (\ref{e46}),
first introduced in the context of quantum mechanics in Ref.~\cite{Z},
suggests a new approach to generating solvable fermionic $\cal PT$-invariant
models whenever there is a boson-fermion duality.

\begin{acknowledgments}
CMB is grateful to the Theoretical Physics Group at Imperial College for its
hospitality and CMB and RJR thank the U.K.~Engineering and Physical Sciences
Research Council for financial support. CMB also thanks the John Simon
Guggenheim Foundation and the U.S.~Department of Energy for financial support.
\end{acknowledgments}


\begin{thebibliography}{99}
\bibitem{SOL}
Another solution is $G_1=i\gamma_5/\delslash$, which again leads to $G_3=G_1/3$,
$G_5=G_1/5$, and so on. However, this solution is unacceptable because it
gives a highly nonlocal expression for $h$.
\end{thebibliography}
\end{document}